\documentclass[lettersize,journal]{IEEEtran}
\usepackage{amsmath,amsfonts}
\usepackage{algorithmic}
\usepackage{algorithm}
\usepackage{array}
\usepackage{booktabs}
\usepackage{makecell}
\usepackage[caption=false,font=normalsize,labelfont=sf,textfont=sf]{subfig}
\usepackage{textcomp}
\usepackage{stfloats}
\usepackage{url}
\usepackage{verbatim}
\usepackage{graphicx}
\usepackage{cite}
\usepackage{todonotes}
\usepackage{xspace}
\usepackage{pifont}
\newcommand{\acronym}{\textit{QPART}\xspace} 

\begin{document}

\title{\acronym: Adaptive Model Quantization and Dynamic Workload Balancing for Accuracy-aware Edge Inference}

\author{Xiangchen Li,~\IEEEmembership{Student Member,~IEEE,}, Saeid Ghafouri, Bo Ji,~\IEEEmembership{Senior Member,~IEEE,}, Hans Vandierendonck,~\IEEEmembership{Senior Member,~IEEE,}, Deepu John,~\IEEEmembership{Senior Member,~IEEE,}, Dimitrios~S. Nikolopoulos,~\IEEEmembership{Fellow,~IEEE,}
\thanks{Manuscript received April 19, 2021; revised August 16, 2021.}}

\markboth{Journal of \LaTeX\ Class Files,~Vol.~14, No.~8, August~2021}%
{Shell \MakeLowercase{\textit{et al.}}: A Sample Article Using IEEEtran.cls for IEEE Journals}


\maketitle

\begin{abstract}
As machine learning inferences increasingly move to edge devices, adapting to diverse computational capabilities, hardware, and memory constraints becomes more critical. Instead of relying on a pre-trained model fixed for all future inference queries across diverse edge devices, we argue that planning an inference pattern with a request-specific model tailored to the device’s computational capacity, accuracy requirements, and time constraints is more cost-efficient and robust to diverse scenarios. To this end, we propose an accuracy-aware and workload-balanced inference system that integrates joint model quantization and inference partitioning. In this approach, the server dynamically responds to inference queries by sending a quantized model and adaptively sharing the inference workload with the device. Meanwhile, the device’s computational power, channel capacity, and accuracy requirements are considered when deciding. 

Furthermore, we introduce a new optimization framework for the inference system, incorporating joint model quantization and partitioning. Our approach optimizes layer-wise quantization bit width and partition points to minimize time consumption and cost while accounting for varying accuracy requirements of tasks through an accuracy degradation metric in our optimization model. To our knowledge, this work represents the first exploration of optimizing quantization layer-wise bit-width in the inference serving system, by introducing theoretical measurement of accuracy degradation. 
Simulation results demonstrate a substantial reduction in overall time and power consumption, with computation payloads decreasing by over 80\% and accuracy degradation kept below 1\%.

\end{abstract}


\section{Introduction}


\IEEEPARstart{A}{lthough} deep learning has had tremendous success in computer vision\cite{krizhevsky2012imagenet}\cite{simonyan2014very}, natural language processing\cite{mikolov2013distributed}\cite{vaswani2017attention} and autonomous transportation\cite{bojarski2016end}, the deployment of these already-trained and frontier models is still performed on server clusters\cite{sze2017efficient}, making the edge users exposed to the limitation of high latency\cite{zhou2019edge}, increased dependency on high transmission bandwidth\cite{shi2016edge}, and potential privacy risks due to data transmission to centralized servers\cite{zhou2019edge}. Many earlier efforts attempted to bring the frontier pre-trained machine learning models to the network's edge. For instance, dedicated hardware accelerators like Google's Edge TPU and Nvidia's Jetson are optimized for performing neural network inference on edge devices\cite{hosseininoorbin2023exploring} \cite{mittal2019survey}. Within the NAS (Network Structure Search) realm, several efforts have tried to discover neural networks optimized for deployment on edge devices\cite{sandler2018mobilenetv2}\cite{tan2019efficientnet}. Model compression is a common and low-cost method to deploy neural networks. Many schemes and frameworks, such as early exit \cite{prechelt2002early}\cite{bakhtiarnia2022single}, model pruning \cite{lin20221xn}\cite{liu2018rethinking}, autoencoders \cite{hu2020fast}, and knowledge distillation \cite{chen2020knowledge}, have been rigorously explored and integrated into the inference offloading problem within the distributed network. From the perspective of computation, inference offloading and model partitioning\cite{kang2017neurosurgeon}\cite{liu2019edge} have been promising tools for reducing the cost of deploying the machine learning models on the edge by offloading partial or all of the computation to more powerful servers. 

While the performance and potential of certain technological advancements facilitating machine learning models on edge devices have been extensively explored in numerous prior studies, their constraints—such as hardware installation costs, model retraining expenses, and the costs associated with fine-tuning—remain significant considerations. Moreover, when deploying machine learning tasks on edge devices, the device type, memory footprint, computation capacity, expected accuracy, and channel conditions impact the inference performance and evolve over time, making it inappropriate to find a universal solution across all future inference queries. Specifically, there are a few factors to be considered before pushing the inference toward the edge of the network: 
\begin{enumerate}
    \item \textbf{Different Hardware Limitations on the Edge:} Edge devices like cell phones, smart watches, network cameras, and VR/AR glasses generate machine learning inference queries but differ in computational and memory capabilities. Different computation capacity comes from the variations in processor architecture, instruction sets, and clock rates \cite{wu2019machine}\cite{bhattacharya2016smart}\cite{zhou2021device}. Devices with powerful processors are more capable of handling inference locally. At the same time, those with low power or limited memory usually use methods like partitioning, quantization, pruning, or distillation to reduce workload. Moreover, the memory capacity also prevents large models to be fully deployed on the edge.
    \item \textbf{Changing Transmission Conditions:} The wireless transmission channel significantly impacts inference service \cite{amiri2020federated}. According to Shannon's capacity formula \cite{shannon1948mathematical}, higher SNR leads to greater bandwidth and lower latency. Hence, higher accuracy may be achieved by trading with latency. Considering channel capacity in inference or offloading makes solutions more robust against network fluctuations.
    \item \textbf{Diverse Accuracy Requirement for Applications:} Desired accuracy varies by application. Critical tasks like medical diagnosis \cite{miotto2018deep}, autonomous vehicles \cite{bojarski2016end}, and power grid monitoring \cite{al2021application} demand high accuracy. In contrast, applications like recommendation systems \cite{hallinan2016recommended} and social media \cite{basarslan2020sentiment} tolerate some accuracy loss. Accepting minor accuracy degradation can often save significant computing resources \cite{han2015deep}\cite{jiao2018energy}. 
\end{enumerate}

\begin{figure}[!t]
    \centering
    \includegraphics[scale=0.33]{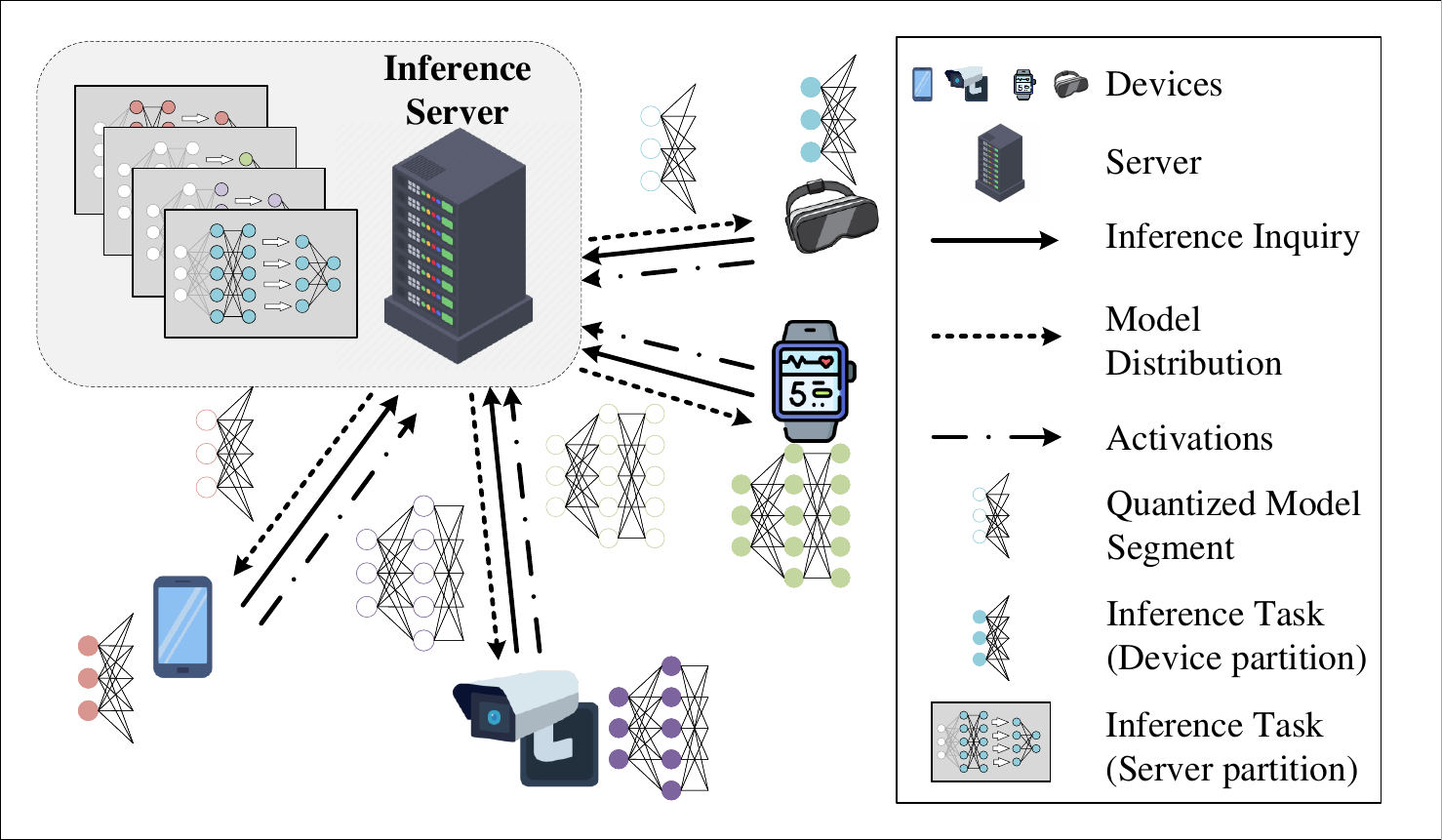}
    \caption{Layer-wise Parameter Size Reduction This figure explains how the \acronym serves the machine learning inference for edge devices. First, an inference request is initialized by edge devices and sent to the server. Second, the server responds to the edge devices by returning a quantized model segment. Then, edge devices process the local inference and feedback on the activation for server inference.}
    \label{fig12}
\end{figure}

To facilitate various devices on the edge with the already-trained neural network models without huge cost, this paper proposes an accuracy-aware and model-less inference system that responds to the inference request from edge devices by automatically quantizing the model for the devices and sharing the computation workload with the device. Shown in Fig.\ref{fig12}, as a machine learning inference infrastructure, our proposed system \acronym maintains the pre-trained machine learning models for a set of edge applications on the cloud server. Once the inference task has been generated from the edge devices, the inference request will be sent to the \acronym specifying the type of the task, the channel capacity and the computation capacity of the device, and the accuracy requirement. Once the \acronym receives the request from edge devices, it solves the joint optimization problem constrained by the accuracy degradation, which is modeled as a layer-wise accuracy degradation measurement, to find the model partition point and layer-wise quantization bit-width according to the current situation. After that, the first part of the whole neural network before the partition point, which is referred as model segment in this paper, will be quantized layer-wise (to reduce the transmission latency and the memory footprint on the device) and then be transmitted from the server to the device for the first stage inference. After the raw data collected by the edge device has been forwarded through the first part of the neural network model on the device, the intermediate output will be transmitted back to the \acronym server for the consecutive inference. Finally, the inference output will be sent back to the device. Our experimental evidence indicates that our framework can achieve up to an 80\%  reduction in communication payload, which contributes the most of the latency in a broad range of edge computing settings\cite{li2019edge}, while limiting accuracy degradation to below 1\%. A comparative analysis with existing model compression methods further validates the efficiency of our approach. 

The contributions of this work are summarized as follows:

\begin{enumerate}
    \item \acronym, an inference serving system consisting of adaptive model quantization and automatic computation workload balancing was proposed. Instead of scaling the pre-trained ML models for the edge manually, inference queries from the edge device are responded by \acronym with optimized model segment for lower latency, energy consumption and cost, also with guaranteed accuracy. 
    \item A joint optimization framework of layer-wise quantization and model partitioning is built for the \acronym system. The computation workload, time consumption, energy consumption, the cost of servers' resources, and transmission latency are modeled within the inference request, forming an optimization problem where the accuracy requirement and memory capacities are modeled as two constraints. 
    \item The proposed optimization problem is well-solved, and the close-form solution of quantization bit-width and partition point are derived. Further, the offline quantization algorithm and online scheduling algorithm are proposed to optimize the partition point and layer-wise quantization in runtime.
    \item By introducing accuracy degradation measurement\cite{zhou2018adaptive} based on quantization noise, the optimization problem is solved, enabling \acronym to optimize quantization bit-width and partition point without violating the accuracy requirement.
    \item A simulation platform is built, and numerous experimental results are carried out to show the efficiency of the proposed \acronym serving inference queries.
    
\end{enumerate}

The paper is organized as follows: Section \ref{sec:backgrounds} reviews the related work of inference offloading and model scaling. Section \ref{sec:system} outlines the target \acronym system settings and the modeling of inference task offloading, including model quantization and partitioning. Section \ref{sec:problem} details the optimization problem's formulation and the corresponding algorithm's introduction. The execution and discussion of simulations are covered in Section \ref{sec:simulations}, and Section \ref{sec:conclusion} provides the conclusion.

\section{Background and Related Work}
\label{sec:backgrounds}
\begin{table}[!t]
\footnotesize
  \caption{\label{tab:comparison}Comparison of related systems accommodating the neural networks on the edge; Latency: Is time consumption considered and optimized?; Energy: Is power consumption considered and optimized?; Accuracy: Is the inference accuracy considered?; Scaling: Does the system scale the model for the edge devices?; Offloading: Does the system support inference offloading from the devices?}
  \centering
  \setlength{\tabcolsep}{2pt}
  \begin{tabular}[]{cccccc}
  \toprule
  System & Latency & Energy & Accuracy &Scaling & Offloading \\
  \midrule
    ApproRR\cite{xu2020energy}& \textcolor{teal}{\ding{51}} & \textcolor{teal}{\ding{51}} & \textcolor{red}{\ding{53}} & \textcolor{red}{\ding{53}} & \textcolor{teal}{\ding{51}} \\
    DeepCOD\cite{yao2020deep}& \textcolor{teal}{\ding{51}} & \textcolor{red}{\ding{53}} & \textcolor{teal}{\ding{51}} & \textcolor{red}{\ding{53}} & \textcolor{teal}{\ding{51}} \\
    DINA-P\cite{mohammed2020distributed}& \textcolor{teal}{\ding{51}} & \textcolor{red}{\ding{53}} & \textcolor{red}{\ding{53}} & \textcolor{red}{\ding{53}} & \textcolor{teal}{\ding{51}} \\
    COB\cite{yang2019computation}& \textcolor{teal}{\ding{51}} & \textcolor{red}{\ding{53}} & \textcolor{red}{\ding{53}} & \textcolor{red}{\ding{53}} & \textcolor{teal}{\ding{51}} \\
    Ly-EXP4\cite{beytur2024optimization}& \textcolor{teal}{\ding{51}} & \textcolor{red}{\ding{53}} & \textcolor{teal}{\ding{51}} & \textcolor{red}{\ding{53}} & \textcolor{teal}{\ding{51}} \\
    MASITO\cite{ben2024multi}& \textcolor{teal}{\ding{51}} & \textcolor{teal}{\ding{51}} & \textcolor{teal}{\ding{51}} & \textcolor{red}{\ding{53}} & \textcolor{teal}{\ding{51}} \\
    INFaaS\cite{romero2021infaas}& \textcolor{red}{\ding{53}} & \textcolor{red}{\ding{53}} & \textcolor{teal}{\ding{51}} & \textcolor{teal}{\ding{51}} & \textcolor{red}{\ding{53}} \\
    FANN-on-MCU\cite{wang2020fann}& \textcolor{red}{\ding{53}} & \textcolor{teal}{\ding{51}} & \textcolor{teal}{\ding{51}} & \textcolor{teal}{\ding{51}} & \textcolor{red}{\ding{53}} \\
    PCA-driven Hybrid-Net\cite{chakraborty2020constructing}& \textcolor{red}{\ding{53}} & \textcolor{teal}{\ding{51}} & \textcolor{red}{\ding{53}} & \textcolor{teal}{\ding{51}} & \textcolor{red}{\ding{53}} \\
    \acronym & \textcolor{teal}{\ding{51}} & \textcolor{teal}{\ding{51}} & \textcolor{teal}{\ding{51}} & \textcolor{teal}{\ding{51}} & \textcolor{teal}{\ding{51}} \\
    
  \bottomrule
  \end{tabular}
\end{table}


As an inference serving system,  \acronym features model quantization and distributed inference, which makes the system located at the intersection of model scaling and inference offloading methods. This section will start by elaborating on these two fundamental techniques that allow the pre-trained model to be deployed on the edge.

\subsection{Deploying Models towards the Edge}

Many modified neural networks or novel learning paradigms have been proposed to address the imbalance between mobile devices and server computational resources. Early exit, model pruning, and auto-encoder, as techniques designed to enhance the computational efficiency of deep neural networks, are beneficial for devices with limited resources. Barbosa et al. \cite{barbosa2022decision} propose the Decision Early Exit (DEEx) approach, which is implemented at a network's first early exit point. DEEx aims to reduce total inference time by avoiding unnecessary evaluations at early exits that do not enhance model performance. Ren et al.\cite{ren2023deep} that the proposed offloading mechanism, aided by model pruning, can significantly reduce latency and energy consumption compared to other algorithms. By proposing a two-step pruning framework for partitioning DNNs between mobile devices and MEC servers, Shi et al.'s work \cite{shi2019improving} effectively reduced the wireless transmission workload of the device or the total computation workload, highlighting the benefits of model pruning in cooperative inference between devices and MEC servers.
Further, auto-encoders were also utilized to facilitate inference on the edge. For instance, Hu et al. \cite{hu2020fast} investigated the streaming inference of CNN to simultaneously improve throughput and accuracy by communication compression. Specifically, an auto-encoder was inserted into the CNN model to compress the communication payload. 

Many machine learning-based inference offloading algorithms effectively reduce latency and maintain accuracy, but they often require modifications to the base model, hindering scalability and adaptability. For instance, early exit mechanisms \cite{li2022graph}\cite{barbosa2022decision}\cite{pacheco2023adaee} necessitate additional training for each exit point, increasing training costs. Model pruning methods \cite{ren2023deep}\cite{shi2019improving} risk accuracy loss when applied across various neural networks, and efforts to mitigate these issues, such as re-tuning, are labor-intensive. Integrating autoencoders \cite{hu2020fast} adds unnecessary computational overhead. In contrast, our proposed model mitigates the computational burden by quantizing model parameters and intermediate outputs to reduced bit widths, thus circumventing additional computations or modifications to the original neural networks.

\subsection{Offloading the Inference Workload on Servers}
Inference offloading, the process of transferring data-intensive and computation-heavy tasks from resource-constrained devices to more capable computing nodes, faces significant bandwidth, latency, and energy consumption challenges. 

Latency is a critical focus in inference offloading. Kang et al. \cite{kang2017neurosurgeon} introduced Neurosurgeon, which reduces inference latency, power consumption, and system throughput by partitioning computation across edge devices and cloud servers. Duan et al. \cite{duan2023optimizing} extended this by jointly optimizing DNN partitioning and pipeline scheduling to lower latency further. Zhao et al. \cite{zhao2020improving} explored the trade-off between latency and accuracy in edge-cloud offloading, focusing on offloading raw data with reduced quality. The challenge of power efficiency in offloading due to limited power on edge devices has also been studied. Xu et al. \cite{xu2020energy} optimized device and server energy consumption in 5G-enabled mobile edge clouds. Sada et al. \cite{sada2024selective} set latency and energy consumption as constraints while optimizing inference accuracy, but their approach required maintaining multiple model variants on the edge, which is impractical. 

In this paper, we proposed \acronym, which, to our best knowledge, is the first inference serving platform that combines automatic model quantization and adaptive inference offloading. By quantizing the model and optimizing the computation offloading pattern on the server, no model variants need to be pre-tuned or stored.

Proposed by Remero et al.\cite{romero2021infaas}, INFaas is an automated model-less inference serving system, which is the closest related work to \acronym. In their work, the model variant searching system \cite{romero2021infaas} responds to the inference queries from a specific platform with an accuracy requirement by searching the optimal neural network variant in the set of scaled neural network models. The developer can specify their application's performance and accuracy requirements without specifying any model-specific parameters or structures. However, INFaaS is only a model variant searching system that is unsuitable for edge devices, and it only tries to search for and respond with the optimal model variant picked from a particular set. Our proposed \acronym system is an end-to-end inference serving platform for edge computing that scales the model and shares the inference workload adaptively. Moreover, to reduce the transmission delay and save the memory footprint on the device, we solve the layer-wise quantization bit-width with close-form solutions instead of manually searching for the optimal model variant within a limited number of fixed model variants pre-defined. Also, NFaaS is not designed for practical edge computing scenarios: It does not consider the device's computation capacity or channel capacity. Table \ref{tab:comparison} provides a comparison summary between \acronym and related works.

\section{System Model}
In this section, we elaborate on our proposed inference serving system \acronym from the perspective of system overview, inference offloading design, model quantization design, and communication model.
\label{sec:system}
\subsection{MEC system}
\begin{figure}[!t]
    \centering
    \includegraphics[scale=0.51]{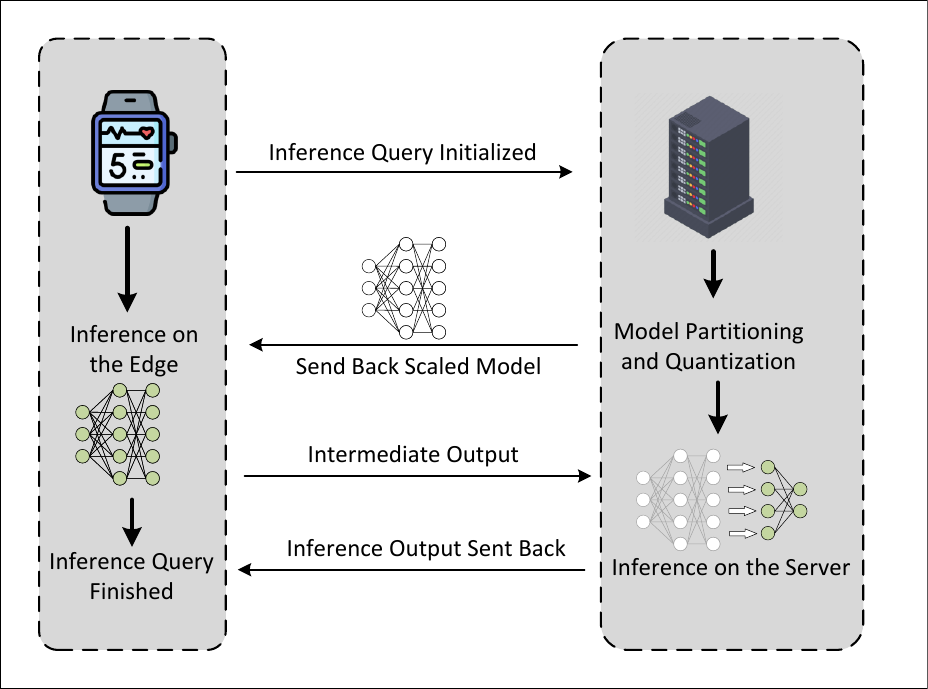}
    \caption{Data Flow between Edge Devices and \acronym System}
    \label{fig1}
\end{figure}
First, we start from the whole picture of our proposed system shown in Fig. \ref{fig1}. Within our proposed inference system, multiple distinct inference requests will be initiated from edge devices, each characterized by varying neural network structures. We denote an inference query $r$ from a mobile device in the time slot $t$ by a tuple $r = (\theta, a)$, where $\theta$ indicates the specific neural network model and $a$ is the accuracy demand, which is defined as the maximal acceptable accuracy degradation compared to the initial accuracy of the model. Once the server receives the inference query, the optimized layer-wise quantization bit-width and model partition point will be solved by the \acronym system. Specifically, model quantization, without degrading the prediction accuracy beyond requirements, effectively decreases the communication payload, thereby reducing communication latency, a primary contributor to overall inference latency in MEC networks \cite{li2019edge}. We split an inference request into two sequential segments through model partitioning, which is processed on the edge device and the server, respectively. Further, the first model segment will be quantized and then sent back to the edge device
. The edge device will perform the inference with a quantized model segment and send the intermediate output to the server for the inference of second model segment
. Once all inference has been operated, the final results will be returned to the device or other nodes if needed.


\subsection{Inference Offloading}
As illustrated in Fig. \ref{fig1}, each inference query $r$ originating from an edge device will be divided into two distinct segments for distributed inference on the device and the server, respectively. We will elaborate on the model partition process and analyze the elements contributing to time and energy consumption, including device-based and server-based inferences and communication-related overhead.

For request $r$ with network model $\theta$, the activation size of layer $l \in \{1,2,\dots, L\}$ is indicated as $z_l^x$, the number of weight parameters in layer $l$ is $z_l^w$. Additionally, we quantify the computational demand of layer $l$ in terms of multiply-accumulate (MAC) operations, referred to as $o(l)$. To adapt our model to various scenarios, we will present the formulation of $o(l)$ for both linear (fully connected) and convolutional layers.
\subsubsection{Linear Layer Computation Cost}
If layer $l$ is a linear layer with input activation $X \in \mathbb{R}^{1 \times D}$ and weight matrix $W \in \mathbb{R}^{D \times G}$, the output activation can be expressed as $Y = X \cdot W, Y \in \mathbb{R}^{1 \times G}$. Hence, the computation cost of layer $l$ can be expressed in Eq. \ref{eq7} \cite{lahmer2022energy}:
\begin{equation}
    o(l) = D \times G \label{eq7}
\end{equation}

\subsubsection{Convolutional Layer Computation Cost}
A standard convolutional layer has the computation cost of\cite{lahmer2022energy}:
\begin{equation}
    o(l) = C_{in} \times C_{out} \times F_1 \times F_2 \times U \times V \label{eq8}
\end{equation}
where $C_{in}$ and $C_{out}$ are the numbers of input channels, and output channels of layer $l$, $F_1$ and $F_2$ define the size of the convolutional filter, and $U$ and $V$ are the size of input activation.

Therefore, with the optimal model partition point given $p$ for request $r$, the computational costs of two segments of the neural network $s$ can be formulated in Eq. \ref{eq9} and Eq. \ref{eq10}:
\begin{equation}
    O_1(p) = \sum_{l = 1}^{p-1} o(l) \label{eq9}
\end{equation}
\begin{equation}
    O_2(p) = \sum_{l = p}^{L} o(l) \label{eq10}
\end{equation}

Let $f^{local} > 0$ denote the clock rate of the edge device when inferring the first part of the request. The time consumption of local inference can be expressed as:
\begin{equation}
    T_{local} = O_1 \cdot \gamma_{local} \cdot \frac{1}{f^{local}} \label{eq11}
\end{equation}
where $\gamma_{local}$ is a constant, referring to the average number of clock cycles needed to finish one MAC operation on the mobile user device. Generally, it depends on the processor architecture, instruction set architecture, pipeline, etc. 

According to \cite{de2014energy}, the power consumption of a processor chip is proportional to $V^2\cdot f_{clock}$, where $V$ represents the circuit supply voltage, and $f_{clock}$ is the clock rate. Furthermore, \cite{burd1996processor} notes that the clock frequency is approximately linearly proportional to the voltage supply, particularly when operating at low voltage limits. Therefore, the energy consumption per clock cycle can be defined as $\kappa f_{clock}^2$, where $\kappa$ denotes the energy efficiency parameter, which depends on the hardware architecture. Consequently, the energy consumption for local inference can be formulated as follows.

\begin{equation}
    E_{local} = \kappa(f^{local})^2 \cdot O_1 \cdot \gamma_{local} \label{eq12}
\end{equation}
where $\kappa_n$ is the energy efficiency parameter of device $n$. In a similar vein, the subsequent expression of time consumption of server-based inference can be obtained:
\begin{equation}
    T_{server} = O_2 \cdot \gamma_{server} \cdot \frac{1}{f^{server}} \label{eq13}
\end{equation}

Inference's energy consumption on the server side is not part of our model due to the data center's continuous power supply. 

As a inference serving platform, \acronym not only tailors the model variant for the edge devices, but also shares the computation workload with the devices, by forwarding the second model segment on the server. The edge devices are supposed to be charged for the computation resources on the server. The cost for occupying the computation resources on the server per second is denoted by $\zeta$. Hence, the overall cost for inferencing the second segment on the server can be defined as Eq.\ref{eq48}:
\begin{equation}
    C = O_2 \cdot \gamma_{server} \cdot \frac{\zeta}{f^{server}} 
    \label{eq48}
\end{equation}

\subsection{Model Quantization}
Given the prevalence of hardware limitations in resource-constrained scenarios, such as embedded systems and wearable devices\cite{qiu2016going}, coupled with the fact that many current neural network models are heavily overparameterized \cite{gholami2021survey}, the focus on training and inference with limited precision has garnered increasing attention among researchers\cite{hubara2017quantized} \cite{zhou2016dorefa}. Model quantization, which involves reducing the precision of weights and biases in machine learning models, especially deep neural networks (DNNs), has emerged as a key technique. This process is integrated into our proposed \acronym because it significantly reduces a neural network's memory footprint, facilitating its deployment on resource-constrained devices and reducing communication latency. 
\begin{figure}[ht]
    \centering
    \includegraphics[scale=0.33]{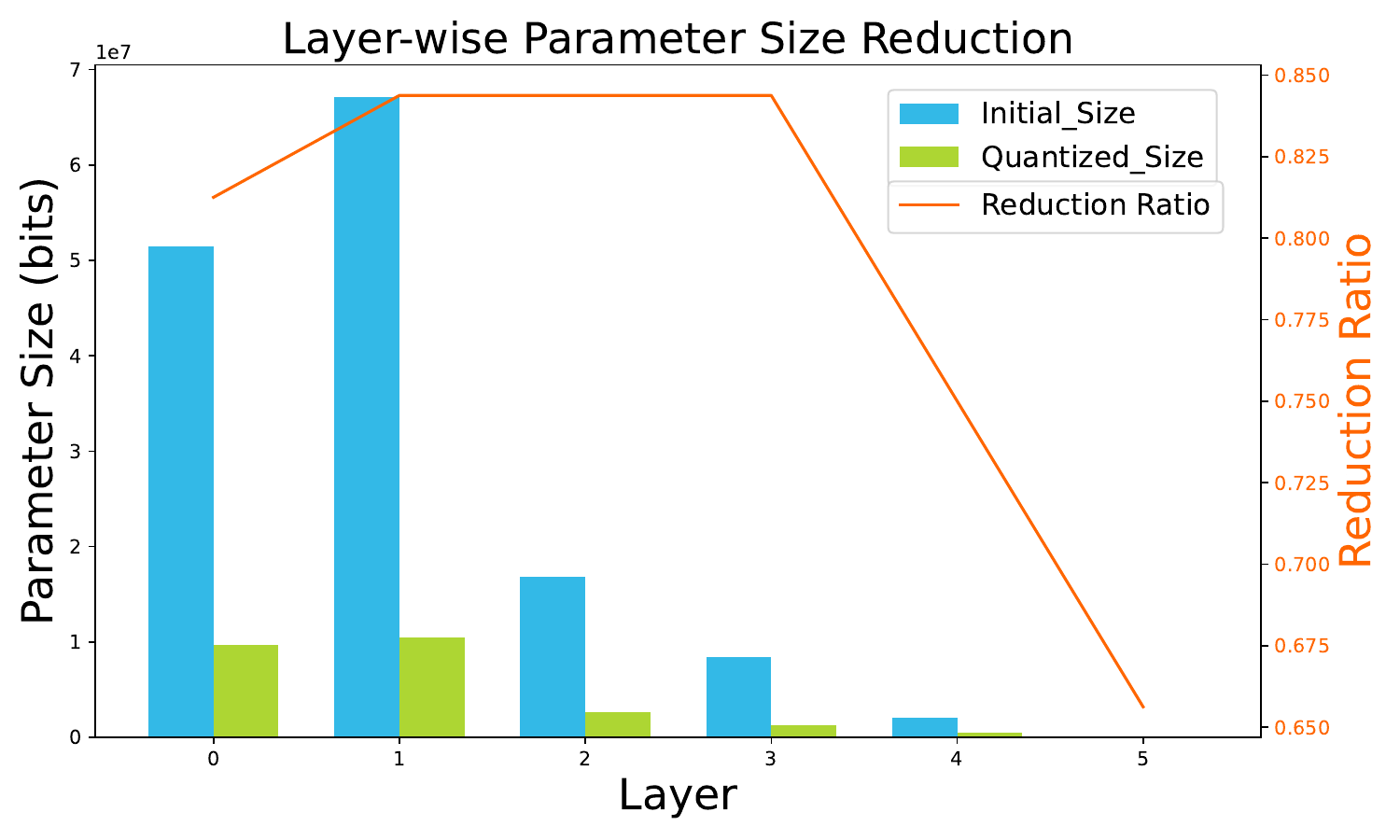}
    \caption{Layer-wise Parameter Size Reduction. From the figure, a substantial part of the model memory footprint can be saved by adopting the quantization method in \acronym}
    \label{fig10}
\end{figure}

Fig. \ref{fig10} shows the layerwise parameter reduction achieved by implementing our proposed algorithm. Across all layers, the sizes of the parameters are substantially reduced by 62\% to 84\%, averaging 77\%, which is referred to the reduction ratio in the figure, while the accuracy degradation remains below 1\%. This figure proves the effectiveness of model quantization in inference offloading. 

We will specifically focus on post-training quantization, which involves quantizing a model after it has been fully trained with full precision. By adopting this approach, our proposed quantized edge inference scheme can be seamlessly integrated into most existing edge inference applications without necessitating the retraining of deep learning models. In the subsequent parts of this section, we will introduce the notations used for quantization and quantized models and activations. 

For an request $r$, we assume that the corresponding neural network model $\theta$ consists of $L$ learnable layers in total and that every layer can be denoted as layer $l, l \in \{1, 2, \cdots, L\}$. The trained model parameters $\theta$ are stored with double floating point precision 
To lower the precision of the model's parameters and the intermediate activation to a more concise format while ensuring that the model's accuracy or generalization capability remains largely unaffected, we first define the uniform asymmetric quantizer as follows:

Given a real value $c$, which can be a weight or activation in neural networks, the quantized value using $b$ bits $c^q$ is selected from the following quantization set $\mathcal{Q}$:

\begin{equation}
    \mathcal{Q} = [\mu:\frac{1}{2^{b}-1}:\phi] + q^z \label{eq1}
\end{equation}
where $\mu$ and $\phi$ define the range of quantization set $\mathcal{Q}$, $[\mu:\frac{1}{2^b-1}:\phi]$ denotes the uniform grid on $[\mu, \phi]$ with interval $\frac{1}{2^b-1}$ between adjacent elements, $q_z$ is the zero point. We define $Q(\cdot)$ as the quantization function, therefore the quantized value $c^q$ is given by Eq.\ref{eq38}:
\begin{equation}
    c^q = Q(c, b, q_z) = \mathop {{\rm{argmin}}}\limits_{{q \in \mathcal{Q}}} (c - q) \label{eq38}
\end{equation}

This paper adopts layer-wise quantization and model partitioning techniques to serve the inference request for edge devices. We employ the notation $p \in {1, 2, \cdots, L}$ to signify the partition point within the model, which serves as a strategic division for processing under our proposed optimization framework. The first neural network segment will be returned to the edge device for local inference. Furthermore, the quantization bit width for each layer in the first neural network segment is defined by a quantization bit-width vector, $\mathbf{b} = [b_i]$, with $i \in {1,2, \cdots, p}$. This vector $\mathbf{b}$ is essential as it determines the quantization granularity for each layer, thereby influencing the model's computational efficiency and overall performance. 

\subsection{Communication Model}
Our proposed MEC system uses a standard wireless communication framework to assess transmission latency. Once the model has been quantized and partitioned for the inference request on the server, the quantized model segment will be transmitted to the edge device through a wireless channel. Also, the intermediate output of the model segment will be transmitted from the device to the server after local inference is completed. Assuming that the transmission time is comparable to the channel coherence period \cite{li2022deep}, the channel gain of the wireless link from the mobile device to server can be defined as Eq.\ref{eq15}:
\begin{equation}
    g = \alpha \cdot h \label{eq15}
\end{equation}
where $h$ is the small-scale fading component, which is frequency-dependent and exponentially distributed with unit mean, and $\alpha$ is the large-scale fading component, which is frequency-independent, reflecting path loss and shadowing. Let $\pi$ denote the transmit power of device, we will have the received SNR shown in Eq. \ref{eq16}:
\begin{equation}
    \beta_k = \frac{\pi \cdot g}{\sigma} \label{eq16}
\end{equation}

Therefore, with a given channel bandwidth $B$, the channel capacity \cite{shannon1948mathematical} can be formulated in the equation. \ref{eq17}:
\begin{equation}
    r = B\cdot log_2(1+\frac{\pi \cdot g}{\sigma}) \label{eq17}
\end{equation}

As for the quantized model parameters and activations to be transmitted, which make up most of the communication payload, we can easily have their data size (in bits) as follows:
\begin{equation}
    Z = Z_x + Z_w = b_p \cdot z_p^x + \sum_{l = 1}^p b_l\cdot z_l^w \label{eq18}
\end{equation}
where $Z_x$ and $Z_w$ are the activation and model parameters communication payload, respectively. Finally, we have the transmission latency $T_{tran}$ and energy consumption $E_{tran}$ of the inference request $r_n^t$ expressed in Eq. \ref{eq19} and Eq. \ref{eq20}:
\begin{equation}
    T_{tran} = \frac{Z}{r} \label{eq19}
\end{equation}
\begin{equation}
    E_{tran} = \pi \cdot \frac{Z}{r} \label{eq20}
\end{equation}

\section{Problem Formulation and Proposed Algorithm}
\label{sec:problem}
In the inference offloading scenario we consider, each user generates inference requests. Our primary objective is to develop an optimal offloading policy that minimizes inference latency and devices' power consumption while ensuring the accuracy of each inference.
Specifically, the optimization problem we aim to solve can be formulated as follows:
\begin{equation}
    \begin{array}{l}
        \mathop {{\rm{min}}}\limits_{{b_i,p_n^t}}\;\;\; \omega \cdot (T_{local}+T_{tran}+T_{server}) +\\ 
        \;\;\;\;\;\;\;\;\;\; \tau \cdot (E_{local} + E_{tran}) + \eta \cdot C \\
        s.t. \;\;\;\; acc_{\theta} - acc_{\theta'} \leq a \\
    \end{array}
    \label{eq21}
\end{equation}
where $\omega, \tau, \eta \geq 0$ are the significance weights for time consumption, energy consumption and cost, respectively. The weights offer the flexibility for each mobile device to tailor its preferences, striking a balance between performances and costs. Further, by letting some weights being zero, the objective can be altered towards different expectations. For example, should a mobile device $n$ exhibit a heightened sensitivity to inference latency coupled with a long battery life, increasing the value of $\omega$ would be beneficial.

Inspired by Yiren\cite{zhou2018adaptive}, we first define quantization noise and adversarial noise to quantify the impact of accuracy requirement $a_n^t$ on the objective function. Specifically, we define $\boldsymbol{\sigma}_l^w$ and $\boldsymbol{\sigma}_p^x$ as the quantization noises on the last activation that are caused by the quantization of $l$-th layer's weights and activation, respectively, and they can be modeled as follows\cite{zhou2018adaptive}: 
\begin{equation}
    ||\boldsymbol{\sigma}_{l}^w||_2^2 = s_l \cdot e^{-ln4 b_l} \label{eq33}
\end{equation}
\begin{equation}
    ||\boldsymbol{\sigma}_{p}^x||_2^2 = s_p \cdot e^{-ln4 b_p} \label{eq34}
\end{equation}

Based on the noises we defined above, two measurements estimating the accuracy degradation caused by quantization of weights and activation in layer $l$ can be defined \cite{zhou2018adaptive}, shown in Eq. \ref{eq23} and Eq. \ref{eq24}:
\begin{equation}
    \psi_l^w = \frac{||\boldsymbol{\sigma}_{l}^w||_2^2}{\rho_l(a)} \label{eq23}
\end{equation}

\begin{equation}
    \psi_l^x = \frac{||\boldsymbol{\sigma}_{l}^x||_2^2}{\rho_l(a)} \label{eq24}
\end{equation}
where $\rho_l(a_n^t)$ is the robustness parameter of layer $l$, and can be calculated with given dataset $\mathcal{D}$ by Eq. \ref{eq25}\cite{zhou2018adaptive}:
\begin{equation}
    \rho_l(a) = \frac{mean(\boldsymbol{\sigma}_l^w, \boldsymbol{\sigma}_l^x)}{mean(\boldsymbol{\sigma^*})} = \frac{\frac{1}{2|\mathcal{D}|} \cdot \sum_{x \in \mathcal{D}}(||\boldsymbol{\sigma}_l^w||_2^2 + ||\boldsymbol{\sigma}_l^x||_2^2)}{\frac{1}{|\mathcal{D}|}\sum_{x \in \mathcal{D}}||\boldsymbol{\sigma}^*||_2^2} \label{eq25}
\end{equation}
where $\boldsymbol{\sigma}^*$ is the adversarial noise on the last layer's activation, which is the minimum noise that can change the classifying result of the neural network. As the measurement of the degradation of the accuracy caused by the quantification in layers, $\psi_l^w = \psi_{l'}^w$ indicates that quantification at layers $l$ and $l'$ affects the accuracy of the model equally. Given that the linearity and additivity of this metric have already been established, and considering the independence of these measurements across all layers, we can reformulate the optimization problem originally presented in Equation \ref{eq21} into a new form, as shown in Eq \ref{eq22}:
\begin{equation}
    \begin{array}{l}
        \mathop {{\rm{min}}}\limits_{{\mathbf{b},p}}\;\;\; \xi \cdot O_1 + \delta \cdot O_2 + \epsilon \cdot (b_p \cdot z_p^x + \sum_{l = p}^L b_l\cdot z_l^w) \\
        s.t. \;\;\;\;  \psi_p^x + \sum_{l=p}^N \psi_l^w \leq \Delta
    \end{array}
    \label{eq22}
\end{equation}
where $\Delta$ is a constant related to model accuracy degradation $a$, specifically, a higher $\Delta$ means a higher $a$. Additionally, local cost $\xi$, server cost $\delta$, and transmission cost $\epsilon$ are defined for simplification, and they can be detailed as follows:
\begin{equation}
    \xi = \frac{\omega \cdot \gamma_{local}}{f^{local}} + \tau\gamma_{local}\kappa_n(f^{local})^2
    \label{eq30}
\end{equation}

\begin{equation}
    \delta = \frac{(\omega+\eta\zeta) \cdot \gamma_{server}}{f^{server}} + \tau\gamma_{server}\eta_m(f^{server})^2
    \label{eq31}
\end{equation}

\begin{equation}
    \epsilon = \frac{\omega + \pi_n \tau}{r_m^n}
    \label{eq32}
\end{equation} 

Then, the optimal value of quantization bit-width $b$ and partition point $p$ can be derived when:
\begin{equation}
\begin{split}
    &\frac{z_p \rho_p}{s_p e^{-ln4b_p}} = \frac{z_{p+1} \rho_{p+1}}{s_{p+1} e^{-ln4b_{p+1}}} = \cdots = \frac{z_{N+1} \rho_{N+1}}{s_{N+1} e^{-ln4b_{N+1}}} \\
    &= \frac{ln4 \cdot [\xi o(p) - \delta o(p) - \epsilon b_p z_p] \rho_p}{s_p e^{-ln4b_p}}
\end{split}
\label{eq29}
\end{equation}
where $z = [z_p^w\; z_{p+1}^w \cdots z_N^w\; z_p^x]$ is defined by connecting parameter size $z_l^w, l \in \{ p, p+1, \cdots , N \}$ and activation size $z_p^x$ . Below is the proof of the Eq.\ref{eq29}.

\textbf{\textit{Proof of Eq.\ref{eq29}}}

In this section, we will prove the optimum of the proposed optimization problem. First, the optimization problem in Eq. \ref{eq22} can be simplified as follows:
\begin{equation}
    \begin{array}{l}
        \mathop {{\rm{min}}}\limits_{{\mathbf{b},p}}\;\;\; \xi \cdot \sum_{l=1}^p o(l) + \delta \cdot \sum_{l=p}^L o(l) + \epsilon \cdot (\sum_{l=p}^{L+1} b_l \cdot z_l) \\
        s.t. \;\;\;\;  \sum_{l=1}^L \frac{s_l e^{-ln4b_l}}{\rho_l} - \Delta \leq 0
    \end{array}
    \label{eq35}
\end{equation}

We denote the objective function as $f(\cdot)$. The following Hessian matrix can check the convexity of the objective function:

\begin{equation}
\begin{array}{l}
H=
\begin{bmatrix}
    \frac{\partial^2 f}{\partial b_p^2}  & \frac{\partial^2 f}{\partial b_p \partial b_{p+1}} & \cdots & \frac{\partial^2 f}{\partial b_p \partial b_{L+1}} & \frac{\partial^2 f}{\partial b_p \partial p}\\
   \frac{\partial^2 f}{\partial b_{p+1} \partial b_p}  & \frac{\partial^2 f}{\partial b_{p+1}^2} & \cdots & \frac{\partial^2 f}{\partial b_{p+1} \partial b_{L+1}} & \frac{\partial^2 f}{\partial b_{p+1} \partial p} \\
   \vdots & \vdots & \ddots & \vdots & \vdots \\
   \frac{\partial^2 f}{\partial b_{L+1} \partial b_{p}} & \frac{\partial^2 f}{\partial b_{L+1} \partial b_{p+1}} & \cdots & \frac{\partial^2 f}{\partial b_{L+1}^2} & \frac{\partial^2 f}{\partial b_{L+1} \partial p} \\
   \frac{\partial^2 f}{\partial p \partial b_{p}} & \frac{\partial^2 f}{\partial p \partial b_{p+1}} & \cdots & \frac{\partial^2 f}{\partial p \partial b_{L+1}} & \frac{\partial^2 f}{\partial p^2}
\end{bmatrix} \\
\\
\;\;\;=
\begin{bmatrix}
    0 & 0 & \cdots & 0 & 0\\
    0 & 0 & \cdots & 0 & 0 \\
   \vdots & \vdots & \ddots & \vdots & \vdots \\
   0 & 0 & \cdots & 0 & 0 \\
   \epsilon z_p & 0 & \cdots & 0 & 0
\end{bmatrix}
\succeq 0
\end{array}
\label{eq36}
\end{equation}

We convert the simplified optimization problem in Eq.\ref{eq35} to Lagrange function below:
\begin{equation}
\begin{array}{cc}
    \mathop {{\rm{min}}}\limits_{{\mathbf{b},p}}\;\;\; \xi \cdot \sum_{l=1}^p o(l) + \delta \cdot \sum_{l=p}^L o(l) + \epsilon \cdot (\sum_{l=p}^{L+1} b_l \cdot z_l) \\
    \;\;\;\;\;\;+ \lambda \cdot (\sum_{l=1}^L \frac{s_l e^{-ln4b_l}}{\rho_l} - \Delta)
    \label{eq46}
\end{array}
\end{equation}

We can solve the gradient vector of function $f(\cdot)$ as follows:
\begin{equation}
     \nabla f = \left[\frac{\partial f}{\partial b_p} \frac{\partial f}{\partial b_{p+1}} \cdots \frac{\partial f}{\partial b_{N+1}} \frac{\partial f}{\partial p} \right] 
     \label{eq39}
\end{equation}
where partial derivative of function $f(\cdot)$ w.r.t. $\mathbf{b}$ and $p$ can be expressed as:
\begin{equation}
    \frac{\partial f}{\partial b_l} = z_l, \;\; l \in \{ p, p+1, \cdots, N+1 \}
    \label{eq40}
\end{equation}
\begin{equation}
    \frac{\partial f}{\partial p} = \xi o(p) - \delta o(p) - \epsilon b_p z_p
    \label{eq41}
\end{equation}

We denote the constrain fuction as $g(\cdot)$, which can be expressed as:
\begin{equation}
    g(\mathbf{b}, p) = \sum_{l=1}^L \frac{s_l e^{-ln4b_l}}{\rho_l} - \Delta
    \label{eq42}
\end{equation}

Similarly, we can derive gradient vector of function $g(\mathbf{b}, p)$ in Eq. \ref{eq43}
\begin{equation}
    \nabla g = \left[\frac{\partial g}{\partial b_p} \frac{\partial g}{\partial b_{p+1}} \cdots \frac{\partial g}{\partial b_{N+1}} \frac{\partial g}{\partial p} \right] 
     \label{eq43}
\end{equation}
where partial derivatives can be detailed as follows:
\begin{equation}
    \frac{\partial g}{\partial b_l} = -\frac{ln4 \cdot s_l}{\rho_l}e^{-ln4 \cdot b_l}, l \in \{p, p+1, \cdots , N+1\}
    \label{eq44}
\end{equation}
\begin{equation}
    \frac{\partial g}{\partial p} = -\frac{s_p \cdot e^{-ln4 b_p}}{\rho_p}
    \label{eq45}
\end{equation}

By using KKT conditions on the Lagrange function defined in Eq.\ref{eq46}, we have the following equations:
\begin{equation}
\begin{split}
     z_p &= \lambda\cdot \frac{ln4\cdot s_p}{\rho_p}e^{-ln4b_p} \\
     z_{p+1} &= \lambda\cdot \frac{ln4\cdot s_{p+1}}{\rho_{p+1}}e^{-ln4b_{p+1}} \\
     &\vdots \\
     z_{N+1} &= \lambda\cdot \frac{ln4\cdot s_{N+1}}{\rho_p}e^{-ln4b_{N+1}} \\
     \xi o(p) - \delta o(p) - \epsilon b_p z_p &= \lambda \cdot \frac{s_p \cdot e^{-ln4 b_p}}{\rho_p}
\end{split}
\end{equation}

Solving these equations, we have:
\begin{equation}
\begin{split}
    &\frac{z_p \rho_p}{s_p e^{-ln4b_p}} = \frac{z_{p+1} \rho_{p+1}}{s_{p+1} e^{-ln4b_{p+1}}} = \cdots = \frac{z_{N+1} \rho_{N+1}}{s_{N+1} e^{-ln4b_{N+1}}} \\
    &= \frac{ln4 \cdot [\xi o(p) - \delta o(p) - \epsilon b_p z_p] \rho_p}{s_p e^{-ln4b_p}}
\end{split}
\label{eq37}
\end{equation}

By further combining the first term and the last term in Eq.\ref{eq37} for resolution, we have:
\begin{equation}
    b_p = \frac{\xi o(p) - \delta o(p) - \frac{z_p}{ln4}}{\epsilon z_p}
    \label{eq47}
\end{equation}
which means we can directly get quantization bit-width for layer $p$ with given partition point $p$. In the rest of this section, we elaborate on the calculation of $\mathbf{b}$ and $p$ by proposing two algorithms according to the solution above.

From the solution in Eq.\ref{eq37} and Eq.\ref{eq47}, we can observe that layer-wise quantization bit-width is solved by iteratively calculating the adversarial noise $\Bar{\sigma^*}$ and robustness parameter $\rho_l$ from the first layer to the last layer of the first model segment. Multiple forwarding operations are involved in this process of calculating adversarial noise, which renders it impractical for real-time deployment. This is a common challenge in many optimization problems, particularly in systems with strict real-time constraints. To address this, many works in the literature resort to approximate or heuristic-based methods that trade off some degree of optimality for practical applicability. In this case, we design the inference serving strategy \acronym with two following algorithms.
\begin{enumerate}
    \item \textbf{Offline Quantization Algorithm: }The optimal layer-wise quantization bit-width is solved by the offline quantization algorithm and it's pre-computed under various system conditions. By optimizing the layer-wise quantization offline, most of the calculation overhead can be dealed in advance and the inference serving system can hence respond to the requests in real-time.
    \item \textbf{Online Serving Algorithm: }With quantization patterns been computed  by offline algorithms, the online serving algorithm responds to the inference request by searching the layer-wise quantization pattern and partition point among pre-calculated results.
\end{enumerate}

\subsection{Offline Quantization Algorithm}
The offline quantization algorithm is shown in Algorithm.\ref{alg:1}, which takes a parameterized model $\theta$ as input, and generates a set of model variants for all possible partition point and several different accuracy level. Specifically, the proposed \acronym enumerate the partition point from the first layer to the last layer $p \in \{1, 2, \cdots, L\}$, with 5 different accuracy degradation requirements $a \in \{a_1, a_2, a_3, a_4, a_5\}$, and optimize the quantization bit-width $\mathbf{b}_a^p$ accordingly. Then the set of quantization pattern $\{(\mathbf{b}_a^p, p)\}_{\theta}$ will be stored for the online serving algorithm in runtime.

\begin{algorithm} 
  \caption{\label{alg:1}Offline Model Quantization Algorithm for Inference offloading}
  \footnotesize{}
  \begin{algorithmic}[1]\normalsize
    \REQUIRE Model $\theta$
    \ENSURE Set of quantized and partitioned model variants $\{\theta_a^p\}$, where $a\in \{a_1, a_2, a_3, a_4, a_5\}, p \in \{1, 2, \cdots, L\}$
    \FOR{$a$ in $\{a_1, a_2, a_3, a_4, a_5\}$}
    \FOR{$p$ = 1, L}
    \FOR{$l$ = $p$, 1}
    \IF{$l == p$}
    \STATE Calculate bit-width at partition point $p$: $b_p = \frac{\xi o(p) - \delta o(p) - \frac{z_i}{ln4}}{\epsilon z_p}$
    \ENDIF
    \STATE Calculate average adversarial noise $\Bar{\sigma^*}$ with given dataset $\mathcal{D}$
    \STATE Incrementally introduce noise into the parameter of layer $l$, observe the accuracy degradation, and record the noise threshold $\sigma_l$ at which the accuracy degradation equals $a$.
    \STATE Calculate robustness parameter $\rho_l$ by Eq.\ref{eq23}
    \STATE Calculate $s_l$ using Eq. \ref{eq33}, by fixing the $b_l$
    \STATE Calculate the quantization bit width $b_l$ for layer $l$, by Eq. \ref{eq29}
    \ENDFOR
    \STATE Save $\mathbf{b}_a^p = (b_1, b_2, \cdots, b_L)$
    \ENDFOR
    \ENDFOR
    \RETURN Set of optimized quantization and partition pattern $\{(\mathbf{b}_a^p, p)\}_{\theta}$ \end{algorithmic}
\end{algorithm}

 For each model quantization pattern $\mathbf{b}_a^p$ with given partition point $p$ and accuracy degradation requirement $a$, to optimize the layer-wise quantization bit width, we start form the layer $p$. The bit width for the layer at the partition point can be directly derived from Eq.\ref{eq29}. Subsequently, the bit width for the preceding layer $b_l, l \in { p-1, p-2, \cdots, 1}$, is solved sequentially from layer $p-1$ to layer $1$ by substituting the parameters of the previous-calculated layer $b_{l+1}$, $\rho_{l+1}$, $s_{l+1}$, $z_{l+1}$, and current layer $\rho_{l}$, $s_{l}$, $z_{l}$ with actual values in Eq.\ref{eq29}. 

\subsection{Online Serving Algorithm}
In the online serving stage, the \acronym responds to the request from the edge devices by searching for the optimal quantization and partitioning pattern $(\mathbf{b}_{a^*}^{p^*}, p^*)_{\theta}$ within the set $\{(\mathbf{b}_a^p, p)\}_{\theta}$, that meets the requirement of the accuracy degradation, and then quantizing the pre-trained model $\theta$, shown in Algorithm.\ref{alg:2}
\begin{algorithm} 
  \caption{\label{alg:2}Online Inference Serving Algorithm for Edge Devices}
  \footnotesize{}
  \begin{algorithmic}[1]\normalsize
    \REQUIRE Inference Request from Edge Device $(\theta, a, r, \pi, \gamma_{local}, f_{local}, \kappa)$
    \ENSURE Optimized Model Variant $\theta^*$

    \STATE Select $a^*$ to be the maximum from set $\{a_1, a_2, a_3, a_4, a_5\}$ without exceeding $a$
    \FOR{$p = 1:L$}
    \STATE Calculate Objective Function Value in Eq.\ref{eq21}, denoted as $j_p$
    \ENDFOR
    \STATE Select $p^*$ to be the value from $\{1, 2, \cdots, L\}$ that minimizes the objective function value
    \STATE Load optimized quantization and partition pattern $(\mathbf{b}_{a^*}^{p^*}, p^*)_{\theta}$ from $\{(\mathbf{b}_a^p, p)\}_{\theta}$
    \STATE Quantize the model $\theta$ to be $\theta^*$ according to Eq.\ref{eq38}
    
    \RETURN $\theta^*$ \end{algorithmic}
\end{algorithm}

\section{Simulations and Numerical Results}
\label{sec:simulations}
\begin{figure}[ht]
    \centering
    \includegraphics[scale=0.75]{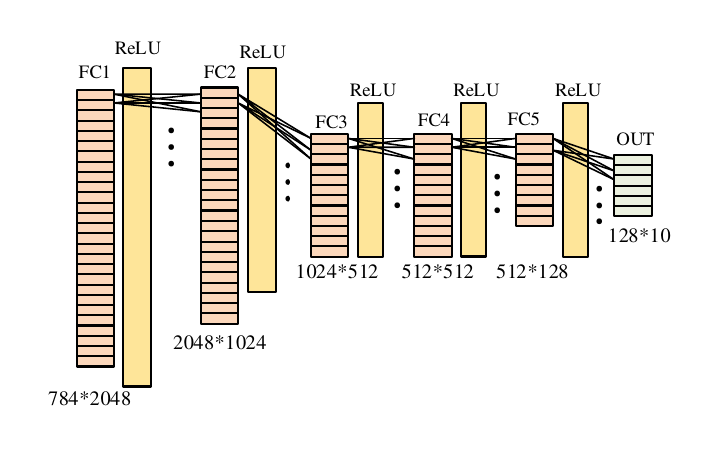}
    \caption{Inference Task Model}
    \label{fig2}
\end{figure}

In this section, we present the numerical results obtained from our simulation platform, which validate the effectiveness of our proposed model quantization and partitioning algorithm for inference offloading. Specifically, we have implemented a DNN-based classifier for the MNIST handwritten digits dataset \cite{lecun1998gradient} to demonstrate the inference offloading capabilities of our approach. This DNN comprises six fully connected layers, as illustrated in Fig.\ref{fig2}. We have meticulously gathered and analyzed the simulation results, including time and energy consumption measurements, overall costs, and accuracy. Furthermore, to gain a more comprehensive perspective on compression-based inference offloading schemes and thoroughly examine the efficacy of our proposed algorithm \acronym in the context of compression-based inference offloading, we have performed comparative tests with auto-encoder-based offloading\cite{yao2020deep}, model-pruning-based offloading\cite{ren2023deep} \cite{shi2019improving}, and direct offloading without the application of any optimization methods. 

We set up the simulator with Python script, within which there are two components: executing module, communication module and performance module. The executing module simulates the processing behaviors of edge devices and server according to the parameters shown in Table \ref{tab:parameters}, and execute the two model segments with different processing profile. The communication module simulates the transmission of the model parameter and intermediate output between the edge device and the server based on the wireless communication model defined in Section \ref{sec:system}. The performance module collects the performance metrics from the other two modules, such as time consumptions and energy consumptions. Specific simulation parameters are summarized in Table \ref{tab:parameters}:




\begin{table*}[!t]
\footnotesize
  \caption{\label{tab:parameters}Simulation Settings}
  \centering
  \begin{tabular}[]{lll}
  \toprule
  Parameters & Interpretation & Value \\
  \midrule
    $\gamma_{local}$& \makecell{Average clock cycles \\needed for a MAC Operation\\ on mobile device} & 5 \\
    $\gamma_{server}$ & \makecell{Average clock cycles \\needed for a MAC \\ Operation on server} & 5/4 \\
    $f^{local}$ & Mobile device clock rate & 200 MHz \\
    $f_{server}$ & Server clock rate &3 GHz \\
    $\pi_n$ & Mobile device transmit power & 1 watt \\
    $r_{n,m}$ & Channel capacity & 200 Mbps \\
    $\omega$ & Time consumption weight & 1 \\
    $\tau$ & Energy consumption weight & 1 \\
    $\kappa_{n}$ & Energy efficiency parameter of mobile device & $3\times10^{-27}$ \\
    $\eta_{m}$ & Energy efficiency parameter of edge server & $3.75\times10^{-27}$ \\
  \bottomrule
  \end{tabular}
\end{table*}

\begin{figure}[ht]
    \centering
    \includegraphics[scale=0.33]{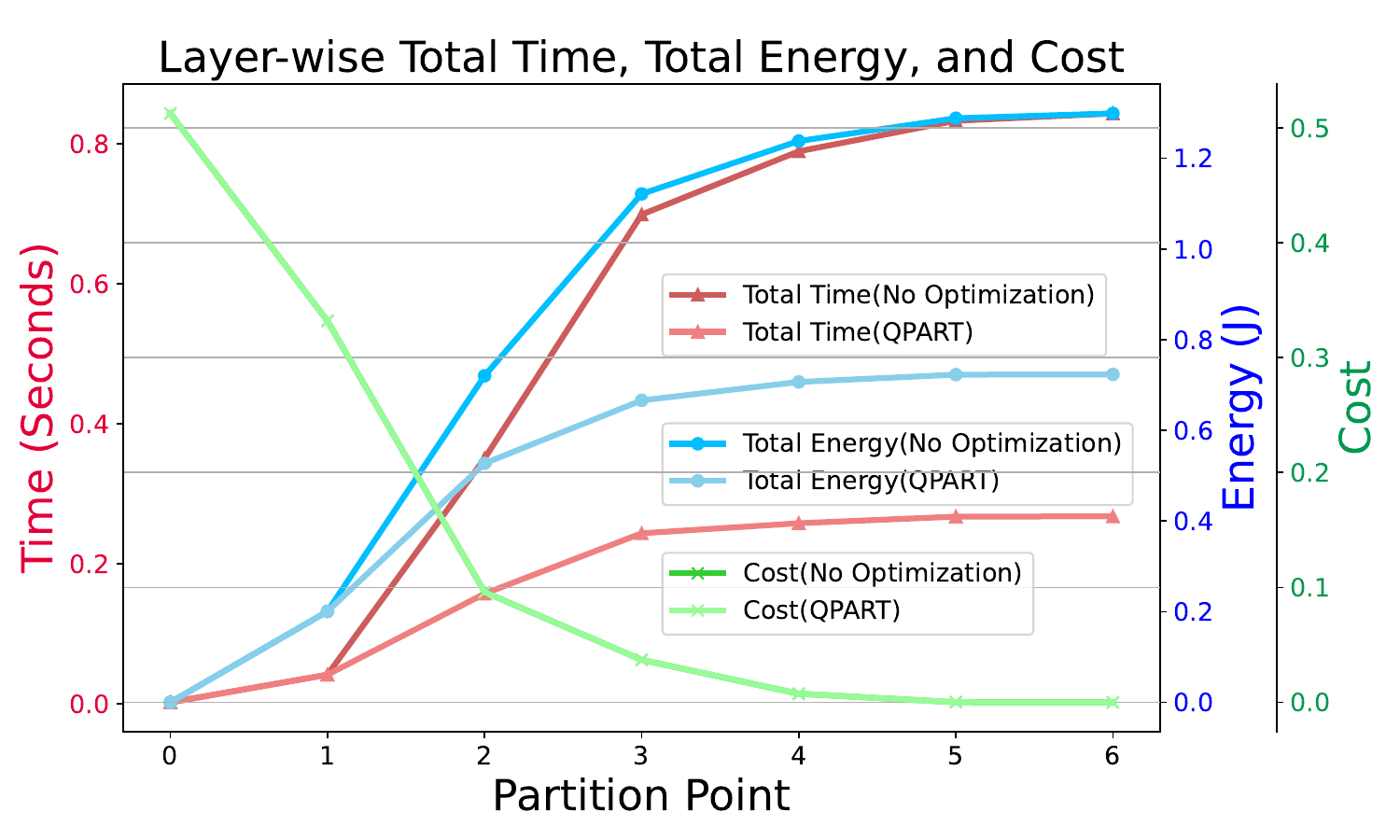}
    \caption{Layer-wise Performance Comparison}
    \label{fig5}
\end{figure}

First we enumerate the model partition point and calculate the quantization bit-width for subsequent layers. As the partition point is incrementally shifted from the first layer towards the final layer, we plot the total time consumption, energy consumption and server cost, which is shown in Fig.\ref{fig5}.

Analysis of Fig.\ref{fig5} reveals a trend: As the partition point moves closer to the last layer, the time consumption and power consumption increase while the server cost decreases. The reason for the increase in time consumption and energy consumption along the partition point moving towards the output layer is that the more model parameters would be counted into the communication payload from the \acronym platform to the device. Meanwhile the server cost is reducing because the shared computation workload on the server decrease while more calculations done in the device. Therefore, a trade-off in computation workload sharing becomes evidet: attempting to save time and energy by allocate more computation on the server increase server cost. Specifically, we compared the layer-wise performance metrics under two strategies, the \acronym and the inference service without optimization, which means the model segments are transmitted from server to the devices without initial pre-trained model intact. From Fig.\ref{fig5}, we can find out the \acronym reduce the power consumption and time consumption significantly in all possible partition layers.

\begin{figure}[ht]
    \centering
    \includegraphics[scale=0.65]{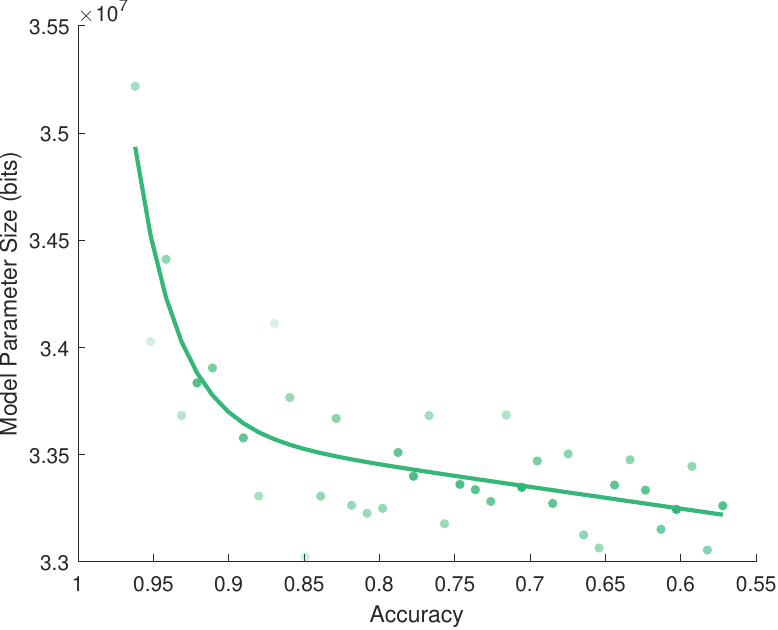}
    \caption{Optimized Model Size v.s. Accuracy}
    \label{fig9}
\end{figure}

By quantizing the model parameter properly, the time consumption and power consumption are hugely reduced, however the lower bit width doesn't come for free. According to Eq.\ref{eq33} and Eq.\ref{eq34}, the quantization noise will increase while the quantization bit width decrease, which will degradate the inference accuracy. Fig.\ref{fig9} illustrates this trade-off between accuracy and model parameter size by establishing various accuracy degradation thresholds and testing the optimized parameter size. This relationship is discerned by aggregating the sizes of the parameters across all layers. It is evident from the figure that the overall model parameter size can decrease exponentially as the required accuracy level is lowered.

\begin{figure}[ht]
    \centering
    \includegraphics[scale=0.33]{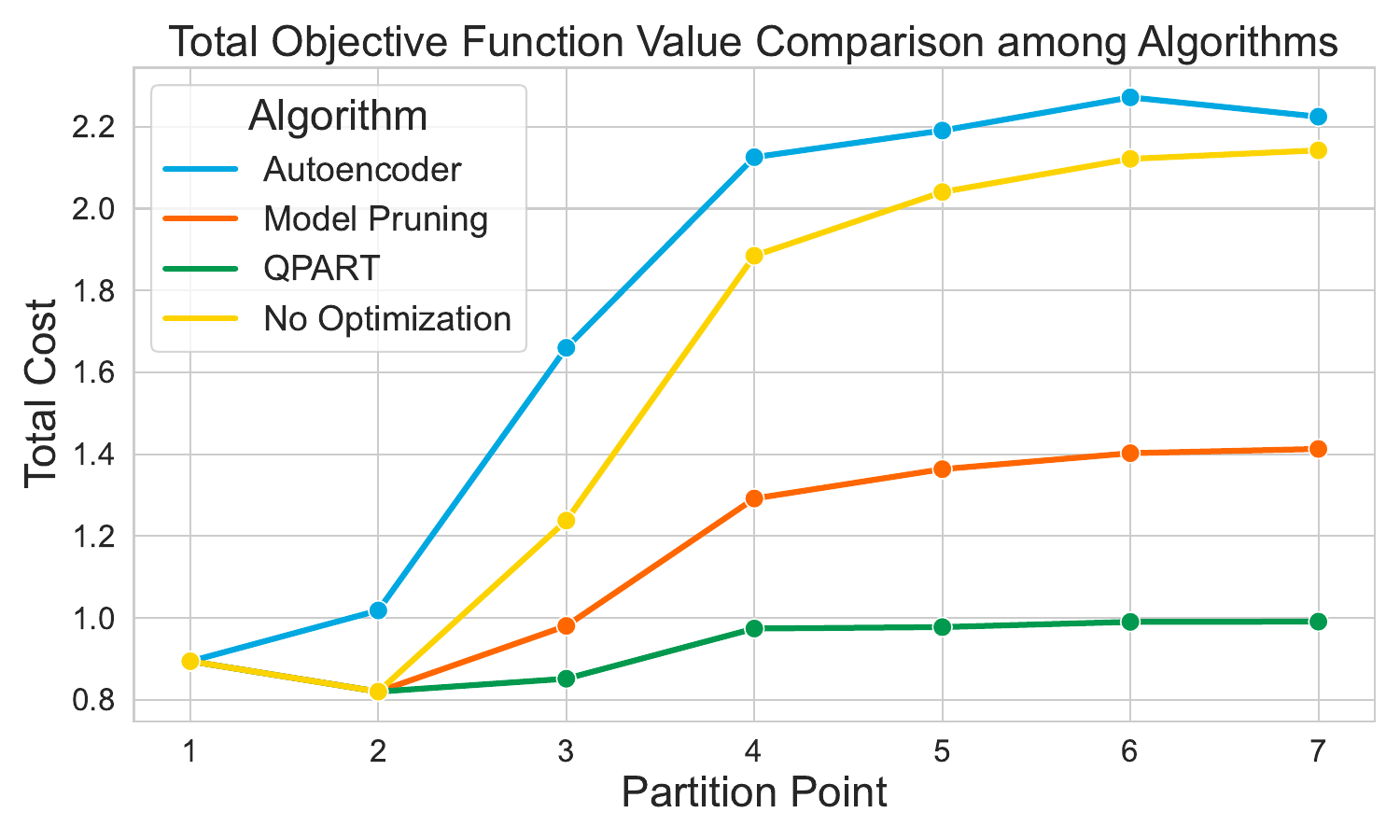}
    \caption{Layer-wise Total Cost Comparison}
    \label{fig8}
\end{figure}

\begin{figure}[ht]
    \centering
    \includegraphics[scale=0.33]{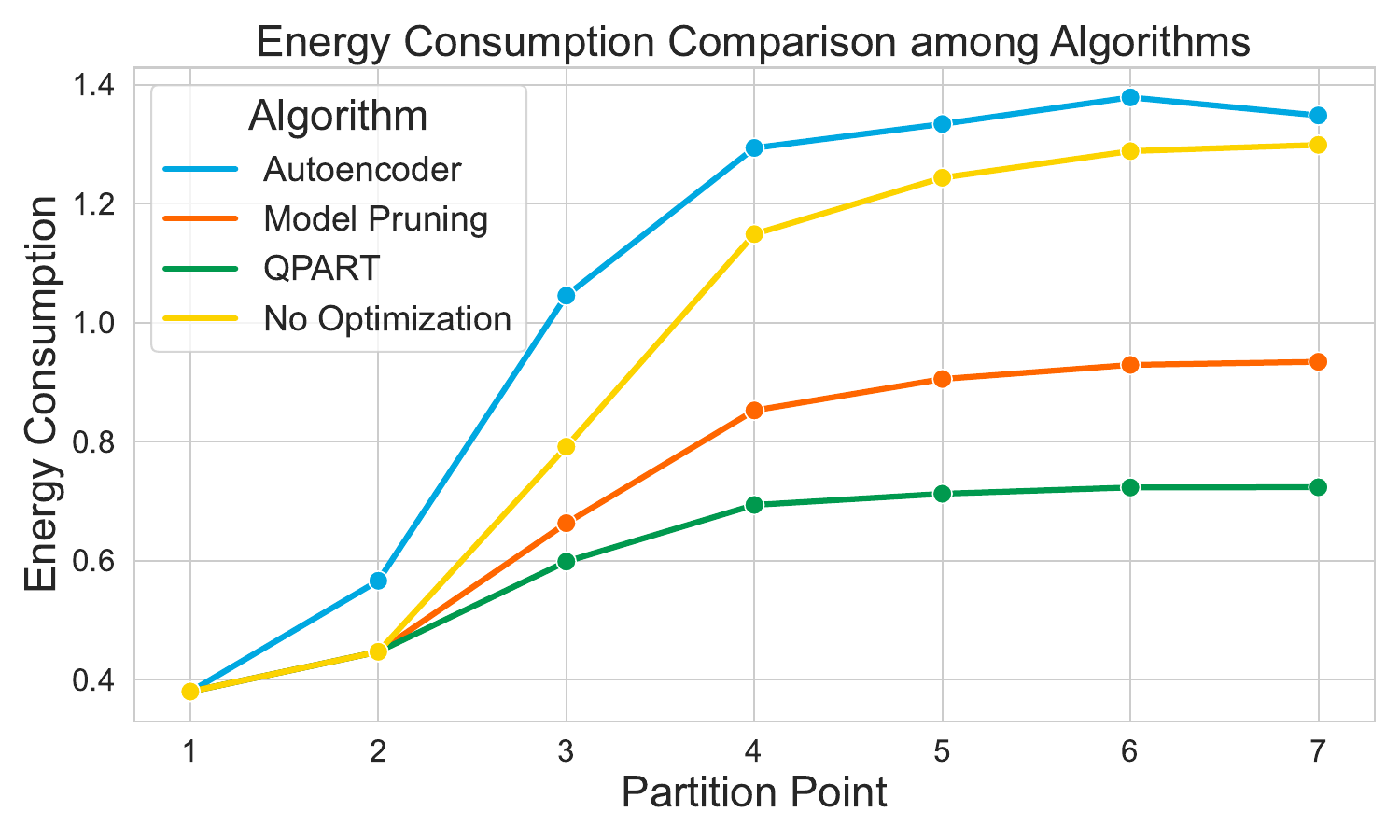}
    \caption{Layer-wise Energy Consumption Comparison}
    \label{fig6}
\end{figure}

\begin{figure}[ht]
    \centering
    \includegraphics[scale=0.33]{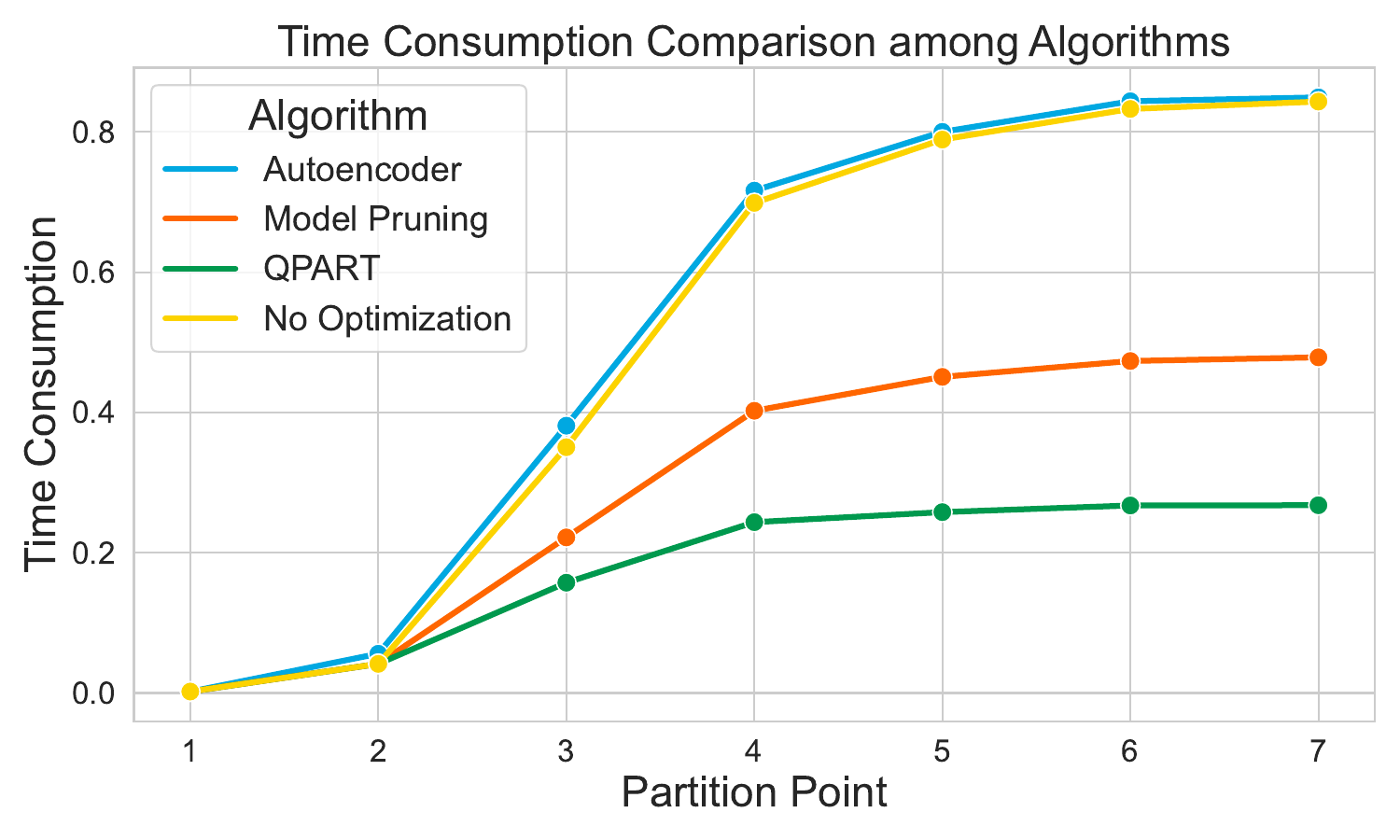}
    \caption{Layer-wise Time consumption Comparison}
    \label{fig7}
\end{figure}

Fig.\ref{fig8}, Fig.\ref{fig6}, and Fig.\ref{fig7} compare the proposed algorithm \acronym with autoencoder-based \cite{yao2020deep} and model pruning-based \cite{ren2023deep} \cite{shi2019improving} inference task offloading algorithms. The auto-encoder-based algorithm compresses the intermediate activation by inserting an encoder and decoder on both the device and the server sides. The model pruning-based method trims neurons in the layers designated for offloading to the server, ensuring that the number of pruned neurons is balanced to maintain accuracy degradation comparable to that of the \acronym algorithm. This strategic pruning of neurons facilitates a reduction in the size of the communication payload. Fig.\ref{fig8} illustrates the objective function values of four offloading schemes, varying according to the partition point. The \acronym algorithm demonstrates the ability to achieve the lowest overall cost. Notably, the autoencoder-based approach tends to incur a significantly higher overall cost than the other three schemes, a finding further substantiated by the subsequent figures.

\begin{figure}[ht]
    \centering
    \includegraphics[scale=0.33]{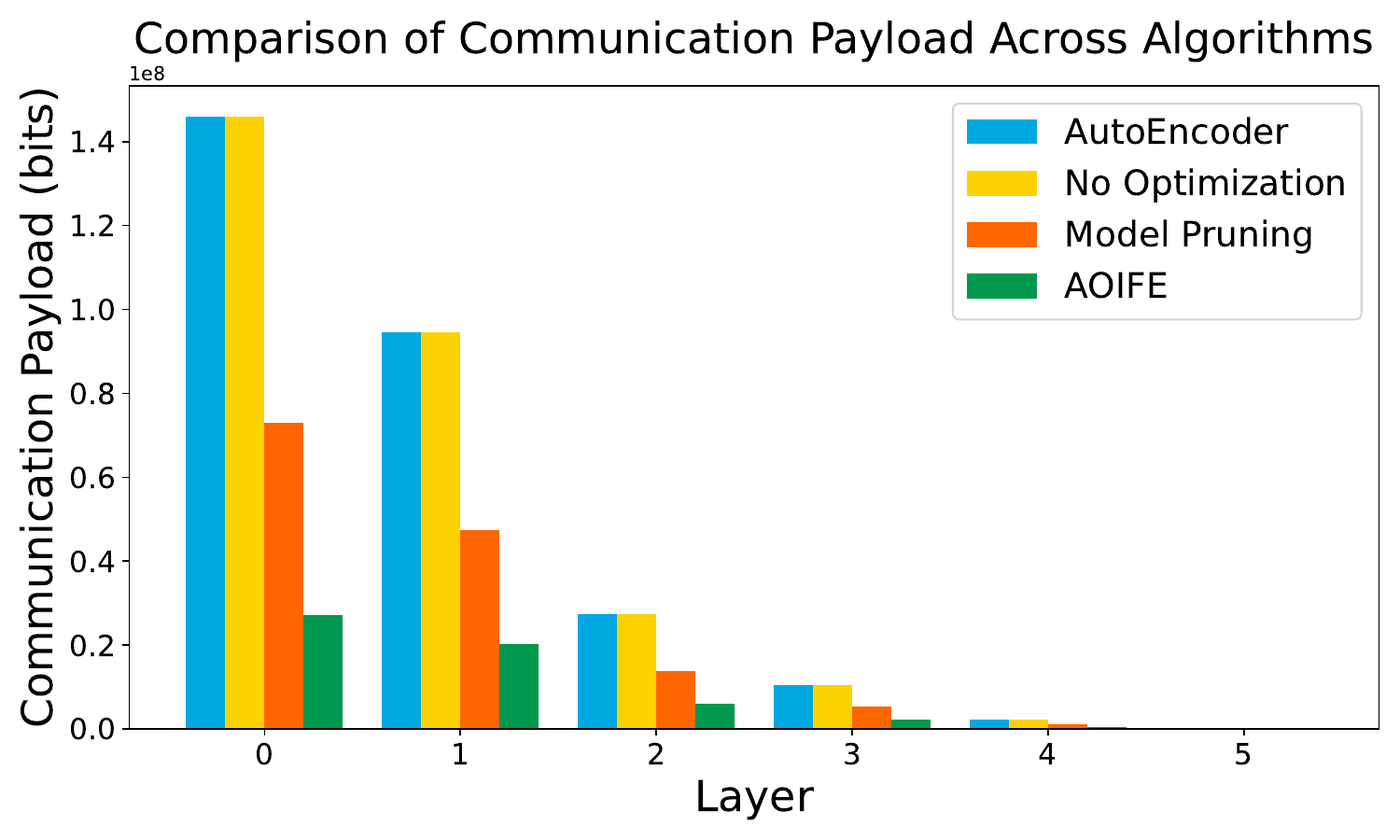}
    \caption{Layer-wise Communication Payload Comparison}
    \label{fig11}
\end{figure}

Fig.\ref{fig6} and Fig.\ref{fig7} provide detailed breakdowns of the time and energy consumption for the four offloading schemes. It is readily apparent that the \acronym algorithm consistently exhibits the lowest energy and time consumption, regardless of the partition point. Due to the incorporation of additional encoding and decoding layers, the auto-encoder-based method incurs greater time consumption and, consequently, higher energy consumption than the other three schemes. Despite this, it slightly reduces communication payload, as shown in Fig.\ref{fig11}.


Table \ref{tab:accuracy} presents the inference accuracy associated with the above four offloading schemes. We evaluated the model's accuracy using the MNIST test dataset, applying four distinct offloading strategies. In scenarios where no optimization algorithm is incorporated into the offloading scheme, the inference accuracy remains consistent with that of the unmodified neural network. The auto-encoder-based method tends to demonstrate improved prediction accuracy when more layers are processed locally. This trend can be attributed to the layers closer to the input being more susceptible to the recovery noise introduced by the autoencoder. The proposed \acronym algorithm maintains accuracy successfully and surpasses the other two compression-based methods in performance.



\begin{table*}[!t]
\footnotesize
  \caption{\label{tab:accuracy}Accuracy of Algorithms at Different Partition Points}
  \centering
  \begin{tabular}[]{lllll}
  \toprule
  Partition Point & Auto-Encoder (\%) & No Optimization (\%) & Model Pruning (\%) & \acronym (\%) \\
  \midrule
    0 & 93.63\% & 96.19\% & 95.03\% & 96.10\% \\
    1 & 93.64\% & 96.19\% & 95.03\% & 96.21\% \\
    2 & 95.91\% & 96.19\% & 95.03\% & 96.21\% \\
    3 & 96.27\% & 96.19\% & 95.03\% & 96.21\% \\
    4 & 95.99\% & 96.19\% & 95.03\% & 96.18\% \\
    5 & 96.15\% & 96.19\% & 95.03\% & 96.20\% \\
  \bottomrule
  \end{tabular}
\end{table*}

\begin{table*}[!t]
\footnotesize
  \caption{\label{tab:baseline}Accuracy on Baseline Models and Datasets}
  \centering
  \begin{tabular}[]{llllll}
  \toprule
    & SVHN & CIFAR10 & CIFAR100 & Resnet18 (ImageNet) & Resnet34 (ImageNet) \\
  \midrule
    Initial Parameter Size (MB) & 5.89 & 46.55 & 46.9 & 107.96 & 180.01 \\
    Optimized Parameter Size (MB) & 0.79 & 5.53 & 6.35 & 19.00 & 32.62 \\
    Communication Payload Compression Ratio & 13.45\% & 11.88\% & 13.53\% & 17.60\% & 18.12\% \\
    Initial Accuracy & 86.88\% & 89.52\% & 67.35\% & 62.75\% & 67.21\% \\
    Accuracy Optimized by AOIFE & 86.79\% & 89.4\% & 67.12\% & 62.09\% & 66.63\% \\
    Accuracy Degradation & 0.08\% & 0.12\% & 0.23\% & 0.66\% & 0.58\% \\
  \bottomrule
  \end{tabular}
\end{table*}

We deployed a convolutional neural network (CNN) on the SVHN, CIFAR10, CIFAR100, and ImageNet datasets to validate our proposed framework across a broader range of datasets and baseline neural networks. In addition, we implemented ResNet architectures, including ResNet18, ResNet34, and ResNet50, on the ImageNet dataset to further benchmark our framework. Table \ref{tab:baseline} illustrates the efficacy of our proposed \acronym when applied across these datasets and models. Notably, we observed a significant reduction in communication payload, calculated by dividing the optimized parameter size by the initial parameter size, ranging from 11.88\% to 18.12\%, with negligible accuracy degradation, ranging from 0.08\% to 0.66\%. These results underscore the versatility and efficiency of our proposed model.

\section{Conclusion}
\label{sec:conclusion}
In this paper, we have proposed \acronym, a neural network inference serving system, emphasizing the development of a flexible, accuracy-aware, and retraining-free tailored for different edge devices. This was achieved by integrating model quantization and model partitioning. We have modeled the time and energy consumption associated with local inference, data transmission, and remote inference. Additionally, our approach includes an accuracy degradation measurement within the joint optimization problem, allowing us to identify the optimal quantization bit-width and partition point while considering accuracy requirements. Our simulation results demonstrate a significant reduction in latency and energy consumption, underscoring the effectiveness of our proposed \acronym algorithm. Looking ahead, our future work will explore global scheduling and allocation strategies. We aim to propose a comprehensive hierarchical optimization framework for inference offloading in mobile edge computing systems, which we anticipate will further enhance the system's overall efficiency and effectiveness in managing complex computational tasks.

\section*{Acknowledgments}

This material is based on work supported by the National Science Foundation under Grants Nos 2315851 and 2106634, a Sony Faculty Innovation Award (Contract AG3ZURVF) and a Cisco Research Award (Contract 878201).

\bibliographystyle{IEEEtran}
\bibliography{reference}

\vfill

\end{document}